\documentclass[prd,aps,showpacs,nofootinbib,showkeywords,eqsecnum]{revtex4-1}

\usepackage{graphicx,color,amsmath,amsxtra}
\usepackage{epsf}
\usepackage{amssymb}
\usepackage{enumerate}
\usepackage{hhline}
\usepackage{array}
\usepackage{tabularx}
\usepackage[unicode]{hyperref}
\usepackage{graphicx}                
\usepackage{epstopdf}

\begin{document}
\pagestyle{myheadings}

\title{Stable small spatial hairs in a power-law {\it k}-inflation model}
\author{Tuan Q. Do }
\email{tuan.doquoc@phenikaa-uni.edu.vn}
\affiliation{Phenikaa Institute for Advanced Study, Phenikaa University, Hanoi 12116, Vietnam}
\affiliation{Faculty of Basic Sciences, Phenikaa University, Hanoi 12116, Vietnam}
\date{\today}

\begin{abstract}
In this paper, we extend our investigation of the validity of the cosmic no-hair conjecture within non-canonical anisotropic inflation. As a result, we are able to figure out an exact Bianchi type I solution to a power-law {\it k}-inflation model in the presence of unusual coupling between scalar and electromagnetic fields as $-f^2(\phi)F_{\mu\nu}F^{\mu\nu}/4$. Furthermore, stability analysis based on the dynamical system method indicates that the obtained solution does admit stable and attractive hairs during an inflationary phase and therefore violates the cosmic no-hair conjecture. Finally, we show that the corresponding tensor-to-scalar ratio of this model turns out to be highly consistent with the observational data of the Planck 2018.

\end{abstract}

%


\maketitle
\section{Introduction} \label{intro}
Cosmological principle, which states that our universe is just simply homogeneous  and isotropic on large scales as described the Friedmann-Lemaitre-Robertson-Walker (FLRW) spacetime, has played a central role in cosmology although it is not straightforward to observationally confirm this principle \cite{Saadeh:2016sak,FLRW}. In fact, many theoretical predictions of the so-called cosmic inflation theory \cite{guth}, which is basically based on the cosmological principle, have been well confirmed by the cosmic microwave background radiations (CMB) observations such as the Wilkinson Microwave Anisotropy Probe satellite (WMAP) ~\cite{WMAP} as well as the  Planck one~\cite{Planck}. Remarkably, among the pioneer inflation models \cite{guth}, the Starobinsky model  involving the $R^2$ correction term has still remained as one of the most favorable models in the light of the Planck observation.

 However, some anomalies of the CMB temperature such as the hemispherical asymmetry and the cold spot, which have been detected  by the WMAP and then by the Planck, cannot be explained within the context of the cosmological principle \cite{Schwarz:2015cma}. In other words, understanding the nature of these exotic features might address a slight modification of the cosmological principle. One possible modification we can think of is using the  Bianchi metrics, which are homogeneous but anisotropic spacetimes, instead of  the FLRW one in order to describe the early universe \cite{bianchi,Pitrou:2008gk}. If the early universe was slightly anisotropic, an important question would be naturally addressed: what would the current state of our universe be ? In other words, would it be slightly anisotropic or completely isotropic ? It is worth noting that some observational evidences have claimed that the current universe might not be isotropic but anisotropic \cite{Colin:2018ghy}. On the theoretical side, the so-called cosmic no-hair conjecture proposed by Hawking and his colleagues few decades ago \cite{GH}  might provide a hint to this important question. Basically, this conjecture states that the late time universe should simply be  homogeneous and isotropic, regardless of initial states, which might be inhomogeneous or/and anisotropic  (a.k.a. spatial hairs). Hence, proving this conjecture is really an important task for physicists and cosmologists. Unfortunately, this is not a straightforward thing. In fact, a complete proof of this conjecture has been a great challenge since the first partial proof, using the energy conditions for the Bianchi spacetimes in the presence of cosmological constant $\Lambda$, by Wald \cite{wald,Starobinsky:1982mr,Barrow:1987ia,Barrow:1984zz,Muller:1989rp,inhomogeneous,Carroll:2017kjo}.
 
It is important to note that the Wald's proof is only limited for homogeneous spacetimes, i.e., the Bianchi spacetimes. One might therefore ask what if the spacetime is inhomogeneous.  It is worth noting that  Starobinsky showed in his seminal paper that the Einstein gravity in the presence of cosmological constant $\Lambda$ admits an inflationary solution, which will approach globally to an isotropic but inhomogeneous state at late time, regarding the hairs as primordial scalar and tensor perturbations \cite{Starobinsky:1982mr}. He also showed that the inhomogeneous and time-independent tensor hairs remain outside of the future event horizon of an observer. This result implies that if the inflationary universe became the isotropic de Sitter state at late time, it would be local, i.e., inside of the future event horizon. Subsequently, he and his colleagues extended this interesting result to a power-law inflation scenario \cite{Muller:1989rp}. They confirmed that a global cosmic no-hair conjecture would also not exist in this case. In other words, the Hawking cosmic no-hair conjecture could be local, i.e., valid inside of the future event horizon. Therefore, a global asymptotic structure of inflationary spacetime outside of the event horizon could be beyond the prediction of the cosmic no-hair conjecture.  Along with the Starobinsky's studies, other works  by other people concerning inhomogeneous background spacetimes, e.g., the so-called Tolman-Bondi spacetime, have been done in Ref. \cite{Barrow:1984zz}. In these models, scale factors of background spacetimes are assumed to be functions of not only time coordinate but also spatial coordinate(s). All these papers have obtained the same result that the late time state of the universe would be locally isotropic and homogeneous. Note that these models have not chosen the approach using the small perturbations of de Sitter spacetime \cite{Starobinsky:1982mr,Muller:1989rp}, which turns out to be technically difficult. Instead, they have preferred using an effective approach \cite{Barrow:1984zz}, in which the background spacetimes are initially assumed to be anisotropic and/or inhomogeneous, to examine the validity of the cosmic no-hair conjecture. It turns out that many follow-up studies have also preferred this effective approach, e.g., see Refs. \cite{Barrow:1987ia,barrow06,kaloper,galileon,kao09,MW0,MW}. In this paper, therefore, we will use the this approach for our analysis.

Along with the partial proofs, some counterexamples to this conjecture have been claimed to exist in Refs. \cite{barrow06,kaloper,galileon,kao09,MW0,MW}. However, many of them have been shown to be unstable against field perturbations \cite{kao09}, except a recent supergravity motivated model proposed by Kanno, Soda, and Watanabe (KSW) \cite{MW0,MW}. In particular, this model has been shown to admit a Bianchi type I metric, which is homogeneous but anisotropic, as its stable and attractive solution during an inflationary phase, due to the existence of unusual coupling between scalar and electromagnetic (vector) fields $-f^2(\phi)F_{\mu\nu}F^{\mu\nu}/4$. This result indicates that the cosmic no-hair conjecture is really broken down in the KSW model. Consequently, many papers have appeared to discuss extensively this interesting model \cite{extensions}.  It is worth noting that once the statistical isotropy of CMB is broken, the scalar power spectrum, i.e. the $TT$ correlations, will be modified accordingly as pointed out in Ref. \cite{ACW}. Consequently, some papers have investigated this issue within the framework of the (canonical) KSW models in the light of the WMAP and Planck  \cite{data}. More interestingly, a smoking gun of anisotropic inflation in the CMB such as the $TB$ and $EB$ correlations, which vanish in isotropic inflation, has been investigated systematically in Refs. \cite{Imprint1,Imprint2,Imprint3}. In addition, primordial gravitational waves in anisotropic inflation have also been studied in Ref. \cite{gws}.  Other cosmological aspects of the KSW mode can be found in recent interesting reviews \cite{SD}. 

It is worth noting that some non-canonical extensions of the KSW model, in which a canonical scalar field has been replaced by non-canonical ones such as the Dirac-Born-Infeld (DBI) \cite{WFK1}, generalized ghost condensate  \cite{ghost-condensed}, supersymmetric Dirac-Born-Infeld (SDBI) \cite{SDBI}, and Galileon fields \cite{G}, have been proposed recently. As a result, the cosmic no-hair conjecture has been shown to be violated in all these non-canonical models. This might confirm the leading role of the unusual coupling $-f^2(\phi)F_{\mu\nu}F^{\mu\nu}/4$ in breaking down the validity of the cosmic no-hair conjecture. Hence, studying this conjecture in other non-canonical scalar field models is essentially important.  Therefore, we would like to examine another scenario, in which a {\it k}-inflation model \cite{k-inflation} is allowed to couple to the KSW model, in this paper to see whether the cosmic no-hair conjecture is violated or not. Note that the CMB imprints of anisotropic inflation of non-canonical scalar field, such as the $TB$ and $EB$ correlations, have been investigated in a recent paper \cite{Do:2020ler}. Once these prediction were confirmed by more sensitive primordial gravitational wave detectors, we would be able to figure out the most viable non-canonical anisotropic model. The present model therefore would be a candidate for this classification. For heuristic reasons, we will investigate the corresponding tensor-to-scalar (power) ratio of the anisotropic power-law $k$-inflation. This can be easily done due to our previous works on the CMB imprints of non-canonical anisotropic inflation, partially done in Ref. \cite{SDBI} and fully done in Ref. \cite{Do:2020ler}. As a result, the corresponding tensor-to-scalar ratio of this model will be shown to be highly consistent with the observational data of the Planck 2018.

As a result, the paper will be organized as follows: (i) A brief introduction of this study has been given in Sec. \ref{intro}. (ii) A basic setup of the proposed model will be shown in Sec. \ref{sec2}. (iii) Then, exact anisotropic solutions will be presented in Sec. \ref{sec3}. (iv) Stability analysis based on the dynamical system method of the obtained solution will be investigated in Sec. \ref{sec4}. (v) The corresponding tensor-to-scalar ratio of this model will be investigated in Sec. \ref{sec5}. (vi) Finally, concluding remarks will be written in Sec. \ref{final}.
\section{Setup} \label{sec2}
 As a result, a general scenario of non-canonical KSW model is given by \cite{ghost-condensed,Do:2020ler},
 \begin{equation} \label{general-action}
S = \int {d^4 } x\sqrt {- g} \left[ {\frac{{R}}
{2}  +P(\phi,X)  - \frac{1}
{4}f^2 \left(\phi\right)F_{\mu \nu } F^{\mu \nu } } \right],
\end{equation}
with $F_{\mu \nu }  \equiv \partial _\mu  A_\nu   - \partial _\nu  A_\mu  $  being the  field strength of the vector field $A_\mu$ used to describe the electromagnetic field. Note that the reduced Planck mass, $M_p$, has been set as one for convenience. In addition, $P(\phi,X)$ is an arbitrary function of scalar field $\phi$ and its kinetic $X\equiv -\partial^\mu\phi \partial_\mu\phi/2$,  which was firstly introduced in the so-called {\it k}-inflation \cite{k-inflation}. It is clear that the KSW model of canonical scalar field is just the simplest case with $P(\phi,X)=X+V(\phi)$. Moreover, if $P(\phi,X)$ takes the following form,
\begin{equation}
P(\phi,X) = \frac{1}
{{{\tilde f}\left( \phi  \right)}}\frac{\gamma-1}{\gamma} -V\left( \phi  \right) ,
\end{equation}
then we will have the DBI extension of the KSW model with  $\gamma  \equiv 1/ \sqrt {1 + {\tilde f} (\phi) \partial _\mu  \phi \partial ^\mu  \phi}$ being the Lorentz factor characterizing the motion of the D3-brane \cite{WFK1}. On the other hand, a supersymmetric DBI extension of the KSW model has been proposed in Ref. \cite{SDBI} with $P(\phi,X)$ being of the following form
\begin{align}
P(\phi,X) &=\frac{1}
{{{\tilde f} \left( \phi  \right)}}\frac{\gamma-1}{\gamma} -\Sigma_0^2 V\left( \phi  \right), \\
\Sigma_0&=\left({\frac{\gamma+1}{2\gamma}}\right)^{\frac{1}{3}}.
\end{align}
It is clear that if we take a limit $\gamma \to 1$, or equivalently ${\tilde f} \to 0$, then both (S)DBI
models will reduce to the canonical KSW model.  In addition, another non-canonical extension of the KSW model has been proposed in Ref. \cite{ghost-condensed} with $P(\phi,X)$ assumed to be the generalized ghost condensate form as
\begin{equation}\label{ghost-cond}
 P(\phi,X) = -X + c \exp \left[n\lambda\phi\right] X^{n+1},
\end{equation}
where $c>0$ and $n \geq1$ are all constants. It is noted again that all these models have been shown to admit counterexamples to the cosmic no-hair conjecture. And the CMB imprints of anisotropic inflation for arbitrary $P(\phi,X)$ have been studied in Ref. \cite{Do:2020ler}. 

In this paper, we would like to seek analytical anisotropic solutions and investigate their stability during an inflationary phase for a specific {\it k}-inflation model \cite{k-inflation} with $P(\phi,X)=K(\phi)X +L(\phi)X^2$ coupled to the KSW model as follows
\begin{equation} \label{action}
S = \int {d^4 } x\sqrt {- g} \left[ {\frac{{R}}
{2} +K(\phi)X +L(\phi)X^2  - \frac{1}
{4}f^2 \left(\phi\right)F_{\mu \nu } F^{\mu \nu } } \right],
\end{equation}
here $K(\phi)$ and $L(\phi)$ are all functions of $\phi$. It is noted that the potential $V(\phi)$ has not been introduced in the above action in a sense that the term $L(\phi)X^2$ could play a similar role as $V(\phi)$ normally does in the slow-roll inflation scenario \cite{k-inflation}.
As a result, the corresponding Einstein field equation can be shown to be 
\begin{equation} \label{Einstein}
R_{\mu\nu}-\frac{1}{2}g_{\mu\nu}R - \left(K +2LX\right)\partial_\mu \phi \partial_\nu \phi - \left(KX +LX^2 -\frac{1}{4}f^2 F^{\rho\sigma}F_{\rho\sigma} \right) g_{\mu\nu} -f^2 F_{\mu\gamma}F_\nu{}^\gamma =0
\end{equation}
along with the corresponding equation of motion of the scalar field $\phi$ given by
\begin{equation} \label{scalar}
\left(K+2LX \right) \square \phi + \partial_\mu \left(K+2L X\right) \partial^\mu \phi = -\partial_\phi K X -\partial_\phi L X^2 +\frac{1}{2}f\partial_\phi f F^{\mu\nu}F_{\mu\nu},
\end{equation}
where $\partial_\phi \equiv \partial/\partial \phi $ and $\square \equiv \frac{1}{\sqrt{-g}} \partial_\mu \left(\sqrt{-g} \partial^\mu \right)$. In addition, the corresponding field equation of the vector field turns out to be
\begin{equation}\label{vector}
\partial_\mu \left[ \sqrt{-g} f^2F^{\mu\nu}\right]=0.
\end{equation}
In order to seek anisotropic solutions to this model, we prefer using the following Bianchi type I metric 
\begin{equation}
ds^2 =-dt^2 +\exp\left[ 2\alpha(t) -4\sigma(t) \right] dx^2 +\exp\left[ 2\alpha(t) +2\sigma(t) \right] \left(dy^2+dz^2 \right)
\end{equation}
along with the compatible vector field $A_\mu$  chosen as  $A_\mu   = \left( {0,A_x \left( t \right),0,0} \right)$ \cite{MW0,MW}.  In addition, the scalar field is assumed to be homogeneous, i.e., $\phi =\phi(t)$. Note that the scale factor $\sigma(t)$ is regarded as a deviation from the spatial isotropy governed by the other scale factor $\alpha(t)$. This means that $\sigma(t)$ should be much smaller than $\alpha(t)$ during an inflationary phase. Non-vanishing $\sigma(t)$ will therefore correspond to the existing of spatial hairs \cite{MW0,MW}. It is important to note  that in this paper as well as in many previous papers on the KSW model \cite{MW0,MW,extensions} the hairs have been regarded as the spatial anisotropies characterised by the scale factor $\sigma(t)$ of the Bianchi type I metric. If the cosmic no-hair conjecture was valid within the KSW model  as well as in its extensions, the corresponding anisotropic solutions should be unstable against field perturbations during an inflationary phase, meaning that they should decay to an isotropic state at late time. Otherwise, the validity of the cosmic no-hair conjecture would be broken down by counterexamples.

As a result,  Eq. (\ref{vector}) can be integrated directly to give a solution,
\begin{equation} \label{eq5}
\dot A_x\left({t}\right)=f^{-2}\left({\phi}\right)\exp\left[{-\alpha-4\sigma}\right]p_A,
\end{equation}
with $\dot A_x \equiv dA_x/dt$ and $p_A$ is a constant of integration ~\cite{MW}. Thanks to this solution, the Einstein field equation can be written explicitly as follows (see the Appendix \ref{app1} for a detailed derivation)
\begin{align}\label{field-eq-1}
\dot\alpha^2 &=\dot\sigma^2 +\frac{K}{6} \dot\phi^2 +\frac{L}{4}\dot\phi^4+\frac{f^{-2}}{6}\exp\left[-4\alpha -4\sigma\right] p_A^2, \\
\label{field-eq-2}
\ddot\alpha &= -3\dot\alpha^2  + \frac{L}{4}\dot\phi^4+\frac{f^{-2}}{6}\exp\left[-4\alpha -4\sigma\right] p_A^2,\\
\label{field-eq-3}
\ddot\sigma &=-3\dot\alpha\dot\sigma +\frac{f^{-2}}{3}\exp\left[-4\alpha -4\sigma\right] p_A^2.
\end{align}
In addition, the corresponding field equation of the scalar field reads
\begin{equation}\label{field-eq-4}
\left(K+3L \dot\phi^2 \right)\ddot\phi =-3 \left(K+L\dot\phi^2 \right)\dot\alpha \dot\phi -\frac{	1}{2}\partial_\phi K \dot\phi^2 -\frac{3}{4}\partial_\phi L \dot\phi^4 +f^{-3}\partial_\phi f \exp\left[-4\alpha -4\sigma\right] p_A^2.
\end{equation}
It straightforward to see that if $K=-1$ and $L=\exp[\lambda \phi]$ we will have the corresponding  dilatonic ghost condensate model \cite{ghost-condensed}. 
\section{Power-law solutions} \label{sec3}
In this section, we would like to figure out analytical solutions to the derived field equations shown above by using the following ansatz such as \cite{MW}
\begin{equation}
\alpha (t)= \zeta \log t; ~\sigma(t) = \eta \log t;~ \phi(t) = \xi \log t +\phi_0
\end{equation}
along with the exponential functions of scalar field given by
\begin{align}
K(\phi)&= k_0 \exp \left[\kappa \phi \right],\\
L(\phi)&=l_0 \exp \left[\lambda \phi \right],\\
f(\phi)&= f_0 \exp \left[\rho \phi \right],
\end{align}
where $\zeta$, $\eta$, $\xi$, $\phi_0$, $k_0$, $l_0$, $f_0$, $\lambda$, $\kappa$, and $\rho$ are all additional parameters.
Given this choice, the scale factors of the Bianchi type I metric will be the power-law function of time as follows
\begin{equation}
ds^2 =-dt^2 +t^{2\zeta-4\eta} dx^2 + t^{2\zeta+2\eta} \left(dy^2+dz^2 \right).
\end{equation}
Hence, an inflationary solution will require that $\zeta -2\eta \gg 1$ and $\zeta+\eta \gg1$. It is noted that $\eta \ll \zeta$ according to the constraint $\sigma(t)\ll\alpha(t) $ as mentioned earlier. Hence, $\zeta \gg 1$ will be required for inflationary solutions. Note that power-law inflation was firstly shown to exist in other scenarios, in which the unusual coupling $-f^2(\phi)F^{\mu\nu}F_{\mu\nu}/4$ was not considered, e.g., see Refs. \cite{Barrow:1987ia,Muller:1989rp,Abbott:1984fp}. More interestingly, the validity of the cosmic no-hair conjecture was also examined in the context of power-law inflation found in some of these models \cite{Barrow:1987ia,Muller:1989rp}. In addition, recent observational constraints of isotropic power-law inflation can be found for example in Ref. \cite{Unnikrishnan:2013vga}.

It turns out that $\dot\alpha ^2 =\zeta^2 t^{-2}$, $\dot\sigma^2 =\eta^2 t^{-2}$, $\ddot\alpha = -\zeta t^{-2}$, and $\ddot\sigma = -\eta t^{-2}$. Hence, in order to have a set of algebraic equations from the above field equations, all terms in these field equations must have $t^{-2}$.  As a result, this requirement  will lead to the following constraints for the field parameters such as
\begin{align}
\kappa & =0, \\
\lambda \xi &=2, \\
\rho \xi +2\zeta +2\eta &=1.
\end{align}
It turns out that the last two constraint equations imply that 
\begin{equation} \label{constraint-1}
\zeta = -\eta -\frac{\rho}{\lambda} +\frac{1}{2}.
\end{equation}
Hence, the requirements $\alpha \gg \sigma$, or equivalently $\zeta \gg \eta$, and $\zeta \gg 1$ for inflationary solutions, lead to two constraints: (i) $\lambda <0$ provided that $\rho >0$ and (ii) $\rho \gg |\lambda|$ such as $\rho/|\lambda| \gg 1$.
As a result, a set of algebraic equations can be defined from the field equations \eqref{field-eq-1}, \eqref{field-eq-2}, \eqref{field-eq-3}, and \eqref{field-eq-4} to be
\begin{align}
\label{field-equation-1}
\zeta^2 &= \eta^2 +\frac{k_0}{6} \xi^2 +\frac{u}{4} \xi^4 +\frac{v}{6},\\
\label{field-equation-2}
-\zeta& = -3\zeta^2 +\frac{u}{4} \xi^4 + \frac{v}{6},\\
\label{field-equation-3}
-\eta &= -3\zeta\eta +\frac{v}{3},\\
\label{field-equation-4}
-\left(k_0+3u \xi^2 \right) \xi &= -3 \left(k_0+u\xi^2 \right) \zeta \xi  -\frac{3\lambda}{4}u \xi^4 +\rho v,
\end{align}
respectively. Here $u$ and $v$ are additional variables defined as follows
\begin{align}
u&= l_0 \exp \left[\lambda \phi_0\right],\\
v&= f_0^{-2}p_A^2 \exp \left[-2\rho \phi_0 \right] .
\end{align}
As a result, $v$ can be defined in terms of $\zeta$ and $\eta$ according to Eq. \eqref{field-equation-3} as 
\begin{equation} \label{field-equation-5}
v= 3\eta \left(3\zeta-1\right).
\end{equation}
Furthermore, $u$ can be defined from Eqs. \eqref{field-equation-2}  to be
\begin{equation}
u= \frac{2}{\xi^4} \left(3\zeta-1\right) \left(2\zeta-\eta \right).
\end{equation}
with the help Eq. \eqref{field-equation-5}.
Given these useful results, solving  either Eq. \eqref{field-equation-1} or Eq. \eqref{field-equation-4} gives us non-trivial solutions of $\zeta$,
\begin{equation}
\zeta=\zeta_\pm = \frac{1}{3}-\frac{ \rho }{3\lambda } \pm\frac{\sqrt{\lambda ^2+4 \lambda  \rho -8 \rho ^2-8 {k_0}}}{6\lambda }.
\end{equation}
Since $\zeta$ should be approximated to be $-\rho/\lambda \gg 1$ according to the constraint equation \eqref{constraint-1}, the suitable solution of $\zeta$ should be $\zeta_-$ rather than $\zeta_+$, i.e., 
\begin{equation}
\zeta =\zeta_- =  \frac{1}{3}-\frac{ \rho }{3\lambda } - \frac{\sqrt{\lambda ^2+4 \lambda  \rho -8 \rho ^2-8 {k_0}}}{6\lambda }.
\end{equation}
Hence, the following $\eta$ turns out to be
\begin{equation}
\eta =\frac{1}{6}- \frac{2\rho}{3\lambda} +\frac{\sqrt{\lambda ^2+4 \lambda  \rho -8 \rho ^2-8 {k_0}}}{6\lambda } .
\end{equation}
As a result, the corresponding anisotropy parameter $\Sigma \equiv \dot\sigma/\dot\alpha$ is given by
\begin{align}
\Sigma =\frac{\eta}{\zeta} = \frac{3\lambda^2 -6\lambda \rho -8k_0 +3 \left(\lambda-2\rho \right) \sqrt{\lambda^2 +4\lambda \rho -8\rho^2 -8k_0}}{3\lambda^2 -12\lambda \rho +12\rho^2 +8k_0}.
\end{align}
It is noted that   $k_0$ has been regarded up to now as a free parameter, in contrast to Ref. \cite{ghost-condensed}, in which $k_0$ was initially fixed to be $-1$.  As a result, the positivity $\lambda ^2+4 \lambda  \rho -8 \rho ^2-8k_0 >0$ puts a constraint on $k_0$ as
\begin{equation} \label{constraint-k0-1}
k_0 < \frac{1}{8} \left(\lambda ^2+4 \lambda  \rho -8 \rho ^2 \right).
\end{equation}
In addition, the positivity of $v$ implies, according to Eq. \eqref{field-equation-5}, that $\eta >0$, provided that $\zeta\gg 1$. As a result, this positivity of $\eta$ leads to the following inequality
\begin{equation}\label{constraint-k0-2}
k_0 > \frac{3}{2} \left(\lambda\rho -2\rho^2 \right).
\end{equation}
On the other hand, we would like to have the following equality
\begin{equation}
\frac{\sqrt{\lambda ^2+4 \lambda  \rho -8 \rho ^2-8k_0 }}{6\lambda } \simeq \frac{2\rho}{3\lambda},
\end{equation}
such that $\zeta \simeq -\rho/\lambda \gg 1$.
Consequently, the corresponding value of $k_0$ should be
\begin{equation}\label{constraint-k0-2}
k_0 \simeq -3\rho^2,
\end{equation}
which does safisty the inequalities \eqref{constraint-k0-1} as well as \eqref{constraint-k0-2}. Note that $u$ is always positive during the inflationary phase.
Consequently, the corresponding approximated value of $\zeta$ and $\eta$ can be defined to be
\begin{align}
\zeta &\simeq \frac{1}{4} -\frac{\rho}{\lambda} \simeq -\frac{\rho}{\lambda} \gg 1,\\
 \eta & \simeq \frac{1}{4} \ll \zeta.
\end{align}
Note again that $\lambda$ has been assumed to be negative definite, in contrast to $\rho$. Hence, the anisotropy $\Sigma/H$ now takes an approximated value during the inflationary phase as
\begin{equation}
\frac{\Sigma}{H}\simeq \frac{\lambda}{\lambda-4\rho} \simeq -\frac{\lambda}{4\rho} \ll 1.
\end{equation}
Indeed, $\Sigma$ should be smaller than one in order to be consistent with the current observations \cite{MW0,MW}. Let us provide here a simple comparison between the present model with the KSW model of canonical scalar field \cite{MW}. In particular, the field parameters will be chosen as $\lambda=\pm 0.1$ (the sign $+$ for the KSW model and $-$ for the present model)  and $\rho=50$ (for both models). Accordingly, it turns out that $(\Sigma/H)_{\rm KSW} \simeq  0.0004$ while $\Sigma/H \simeq 0.0005$. Hence, the anisotropy in the present model can be said to be similar to that of the KSW model.
\section{Stability analysis} \label{sec4}
So far, we have found the power-law Bianchi type I solution to the {\it k}-inflation model \cite{k-inflation} in the presence of the unusual coupling between the scalar and electromagnetic fields, i.e., $-f^2(\phi)F_{\mu\nu}F^{\mu\nu}/4$. In this section, we would like to investigate the stability of this anisotropic solution during the inflationary phase in order to see whether the cosmic no-hair conjecture is violated or not. Following the previous studies \cite{MW,WFK1,ghost-condensed,SDBI,G}, the dynamical system method will be used to do this task. Note that there is another stability analysis approach based on power-law perturbations, i.e., $\delta \alpha =A_\alpha t^n$, $\delta \sigma =A_\alpha t^n$, and $\delta \phi =A_\phi t^n$, which would lead to the same conclusion about the stability of the anisotropic solutions \cite{WFK1,SDBI,G}. However, this method does not yield an information of attractive property of the anisotropic solutions.

As a result, introducing dynamical variables such as \cite{MW,WFK1,ghost-condensed,SDBI,G}
\begin{equation}
x = \frac{\dot\sigma}{\dot\alpha};~y= \frac{\dot\phi}{\dot\alpha};~z=\frac{f^{-1}}{\dot\alpha}\exp[-2\alpha -2\sigma]p_A
\end{equation}
along with two auxiliary variables \cite{G}
\begin{equation}
w_\kappa = \exp\left[ \frac{\kappa}{2} \phi \right]; ~w_\lambda = \sqrt{l_0}\dot\alpha \exp \left[\frac{\lambda}{2} \phi \right].
\end{equation}
It is apparent that
\begin{align}
\frac{dx}{d\alpha}& = \frac{\ddot\sigma}{\dot\alpha^2} - \frac{\ddot\alpha}{\dot\alpha^2}x,\\
\frac{dy}{d\alpha}&= \frac{\ddot\phi}{\dot\alpha^2} -\frac{\ddot\alpha}{\dot\alpha^2}y,\\
\frac{dz}{d\alpha}&=- z\left[ 2(x+1) +\rho y \right] -\frac{\ddot\alpha}{\dot\alpha^2} z,\\
\frac{dw_\lambda}{d\alpha}&=  \frac{\lambda}{2} yw_\lambda +\frac{\ddot\alpha}{\dot\alpha^2}w_\lambda,\\
\frac{dw_\kappa}{d\alpha}&= \frac{\kappa}{2} y w_\kappa.
\end{align}
As a result, the field equations \eqref{field-eq-2}, \eqref{field-eq-3}, and \eqref{field-eq-4} can be converted into the autonomous equations of the dynamical variables $x$, $y$, and $z$ as follows
\begin{align} \label{dyn-1}
\frac{dx}{d\alpha} =&~ x\left[x^2+\frac{k_0}{6}y^2w_\kappa^2-1\right] +\frac{z^2}{3},\\
\label{dyn-2}
\frac{dy}{d\alpha}=& -\frac{1}{k_0 w_\kappa^2+3y^2 w_\lambda^2} \left[3y \left(k_0w_\kappa^2+ y^2 w_\lambda^2\right) +\frac{\kappa}{2} y^2 w_\kappa^2-3\lambda \left(x^2 +\frac{k_0}{6}y^2w_\kappa^2 +\frac{z^2}{6}-1\right) -\rho z^2 \right]\nonumber\\
&+y\left[x^2 +\frac{k_0}{6}y^2w_\kappa^2+2 \right],\\
\label{dyn-3}
\frac{dz}{d\alpha}=&~ z \left[ x^2 +\frac{k_0}{6}y^2w_\kappa^2 -2x -\rho y \right] ,
\end{align}
along with that of two auxiliary variables,
\begin{align}
\label{dyn-4}
\frac{dw_\lambda}{d\alpha}=&- w_\lambda \left[x^2 +\frac{k_0}{6}y^2w_\kappa^2-\frac{\lambda}{2} y+2 \right],\\
\label{dyn-5}
\frac{dw_\kappa}{d\alpha}=&~ \frac{\kappa}{2} y w_\kappa.
\end{align}
Here $d\alpha = \dot\alpha dt $ can be understood as a new time coordinate \cite{MW,WFK1,ghost-condensed,SDBI,G}.  In addition, the following constraint equation coming from the Friedmann equation \eqref{field-eq-1},
\begin{equation} \label{dyn-5}
\frac{1}{4} y^4w_\lambda^2 = -x^2 -\frac{k_0}{6}y^2 w_\kappa^2-\frac{z^2}{6}+1,
\end{equation}
has been used to derive the above autonomous equations. Now, we would like to see whether this dynamical system admits anisotropic fixed points with $x\neq 0$. Mathematically, these fixed points are solutions of the following equations, $dx/d\alpha =dy/d\alpha=dz/d\alpha =dw_\lambda/d\alpha=dw_\kappa /d\alpha =0$.  It is noted that  the equation $dw_\kappa /d\alpha =0$ implies the result that $\kappa =0$, or equivalently $w_\kappa =1$. In addition, the equation  $dw_\lambda/d\alpha=0$ implies that
\begin{equation}
x^2 +\frac{k_0}{6}y^2-\frac{\lambda}{2} y+2 =0,
\end{equation}
while the equation $dz/d\alpha=0$ leads to another relation
\begin{equation}
x^2 +\frac{k_0}{6}y^2 -2x -\rho y =0 .
\end{equation}
Here, an isotropic fixed point solution with $x=z=w_\lambda=0$ is not our current interest.
Hence,  a relation between $x$ and $y$ can be figured out from these two equations  as
\begin{equation} \label{def.y}
y=\frac{4(x+1)}{\lambda-2\rho}.
\end{equation}
On the other hand, we can obtain from two equations $dx/d\alpha=0$ and  $dz/d\alpha=0$ a relation as
\begin{equation} \label{def.z}
z^2 =-3x\left(2x+\rho y -1 \right).
\end{equation}
Thanks to these useful relations, both the equations $dy/d\alpha=0$ or $dz/d\alpha=0$, lead to a non-trivial equation for anisotropic fixed points $x\neq 0$ as
\begin{equation}
\left(3\lambda^2 -12\lambda \rho +12\rho^2 +8k_0 \right)x^2 - 2 \left(3\lambda^2 -6\lambda\rho-8k_0 \right)x + 4 \left(6\rho^2 -3\lambda\rho +2k_0 \right) =0.
\end{equation}
As a result, this equation admits two solutions of $x$,
\begin{equation}
x= x_\pm =  \frac{3\lambda^2 -6\lambda\rho-8k_0 \pm 3 \left(\lambda-2\rho\right)\sqrt{\lambda^2 +4\lambda \rho  -8\rho^2 -8k_0 }}{3\lambda^2 -12\lambda \rho +12\rho^2 +8k_0}.
\end{equation}
Comparing these solutions with the power-law solutions obtained in the previous section implies that only the solution,
\begin{equation}
x=x_+
\end{equation}
 is a suitable solution. Indeed, it is straightforward to see that $x_+ =\Sigma/H$, which has been clearly defined in the previous section for the anisotropic power-law solution. And the corresponding value of the other dynamical variables $y$, $z$, and $w_\lambda$ can be defined according to Eqs. \eqref{def.y}, \eqref{def.z}, and \eqref{dyn-5}, respectively. These results indicate that the anisotropic fixed point $x=x_+$ is indeed equivalent to the anisotropic power-law solution derived in the previous section. Consequently, the anisotropic fixed point and the anisotropic power-law solution share the same stability property. Hence, we will investigate the stability of the anisotropic fixed point during the inflationary phase from now on. 
 
\begin{figure}[hbtp] 
\begin{center}
{\includegraphics[height=80mm]{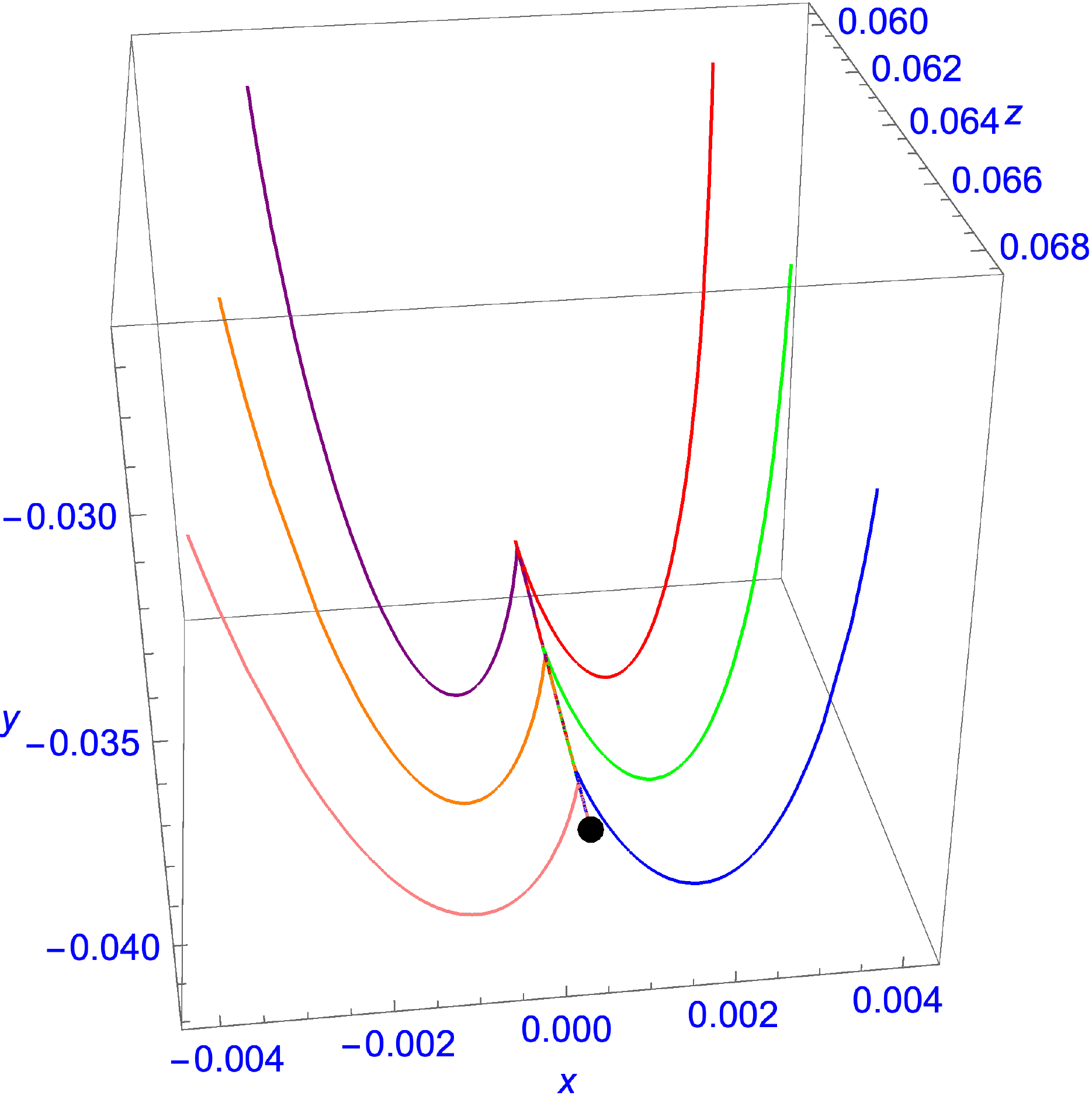}}\\
\caption{Attractor property of the anisotropic fixed point with the field parameters chosen as $\lambda=-0.1$ and $\rho =50$. This plot clearly shows that trajectories with different colors corresponding to initial conditions  all converge to  the anisotropic fixed point (the black point). }
\label{fig1}
\end{center}
\end{figure}

 As discussed above, we have the following constraints as $k_0\simeq -3\rho^2$ and $|\lambda| \ll \rho$ such that $\zeta \simeq \rho/|\lambda| \gg 1$ for the negative $\lambda$ and positive $\rho$ during the inflationary phase. Consequently, we are able to show the corresponding approximated value of the anisotropic fixed point as
 \begin{align}
 x &\simeq -\frac{\lambda}{4\rho} \ll 1,\\
  y &\simeq -\frac{2}{\rho}<0,\\
  z^2 &\simeq 9x = -\frac{9\lambda}{4 \rho} >0,\\
  w_\lambda^2 &\simeq \frac{3\rho^4}{4} \left(\frac{\lambda}{8\rho}+1 \right)>0.
 \end{align}
 To see how small or large this fixed point is, we choose for example $\lambda =-0.1$ and $\rho =50$. As a result, the corresponding anisotropic fixed point will be $(x,~y,~z,~w_\lambda)\simeq (0.0005, ~-0.04, ~0.067,~ 2165)$.
 Now, we would like to perturb the autonomous equations \eqref{dyn-1}, \eqref{dyn-2}, \eqref{dyn-3}, and \eqref{dyn-4} around the anisotropic fixed point. As a result, a set of the following perturbation equations turns out to be
 \begin{align}
 \frac{d\delta x}{d\alpha} \simeq & -3\delta x -\frac{\lambda}{2} \delta y +\sqrt{\frac{-\lambda}{\rho}}  \delta z,\\
   \frac{d\delta y}{d\alpha} \simeq &~\frac{\lambda}{\rho^2}\delta x+ \left(\frac{9\lambda}{4\rho}-7 \right)\delta y +\frac{1}{2} \sqrt{-\lambda\rho} \delta z -\frac{\sqrt{3}}{\rho^2} \left(3\lambda -\frac{4}{\rho} \right)\delta w_\lambda, \\
     \frac{d\delta z}{d\alpha} \simeq & -3 \sqrt{\frac{-\lambda}{\rho}} \delta x + \frac{3}{2} \sqrt{-\lambda\rho} \delta y +\frac{\lambda}{2\rho}\delta z, \\
       \frac{d\delta w_\lambda}{d\alpha} \simeq &~ \frac{\sqrt{3}}{4}\lambda \rho \delta x -\sqrt{3} \rho^3 \delta y- \frac{\lambda}{\rho} \delta w_\lambda.
   \end{align}
   Here we have only considered the leading terms in these perturbation equations for simplicity. 
  Following Ref.  \cite{MW}, we will take the exponential perturbations of dynamical variables given by
     \begin{align}
   \delta x &=A_x \exp[\omega\alpha],\\
  \delta y& =A_y \exp[\omega\alpha],\\
   \delta z &=A_z \exp[\omega\alpha],\\
   \delta w_\lambda& =A_w \exp[\omega\alpha],
   \end{align}
   here the sign of $\omega$ will determine the stability of the anisotropic fixed point. In particular, if a set of perturbation equations admits any positive $\omega$, then $\exp[\omega\alpha]$ will blow up as $\alpha \to +\infty$, causing an instability of the anisotropic fixed point. In contrast, the anisotropic fixed point will be stable if all solutions of $\omega$ turn out to be negative because $\exp[\omega\alpha]$ will tend to zero as $\alpha \to + \infty$. As a result, we are able to obtain the following set perturbation equations, which can be written as a matrix equation as follows
   \begin{equation} \label{stability-equation}
{\cal M}\left( {\begin{array}{*{20}c}
   A_x  \\
   A_{y}  \\
   A_z  \\
   A_{w}\\
 \end{array} } \right) \equiv \left[ {\begin{array}{*{20}c}
   {-\omega-3} & {-\frac{\lambda}{2}} & {\sqrt{\frac{-\lambda}{\rho}}  } & {0 }   \\
   {\frac{\lambda}{\rho^2} } & {-\omega-7+\frac{9\lambda}{4\rho}} & {\frac{1}{2} \sqrt{-\lambda\rho} } & {-\frac{\sqrt{3}}{\rho^2} \left(3\lambda -\frac{4}{\rho} \right)}  \\
   {-3 \sqrt{\frac{-\lambda}{\rho}}} & {\frac{3}{2} \sqrt{-\lambda\rho} } & {-\omega+\frac{\lambda}{2\rho} } & {0} \\
   {\frac{\sqrt{3}}{4}\lambda \rho}&{-\sqrt{3} \rho^3}&{0} &{-\omega- \frac{\lambda}{\rho}} \\
 \end{array} } \right]\left( {\begin{array}{*{20}c}
   A_x  \\
   A_{y}  \\
   A_z  \\
   A_{w}\\
 \end{array} } \right) = 0,
\end{equation}
Mathematically, Eq. \eqref{stability-equation} admits non-trivial solutions if and only if
\begin{equation}
\det {\cal M}=0,
\end{equation}
which can be defined to be a polynomial equation of $\omega$ as follows
   \begin{equation} \label{eq-omega}
 a_4 \omega^4 +a_3\omega^3 +a_2\omega^2 +a_1\omega +a_0 =0,
   \end{equation}
   where 
   \begin{align}
   a_4&=1>0,\nonumber\\
   a_3 &\simeq 10 >0,\nonumber\\
   a_2& \simeq 33-\frac{33}{4}\lambda\rho >0,\nonumber\\
   a_1& \simeq 36 -\frac{99}{4}\lambda\rho >0,\nonumber\\
   a_0& \simeq \frac{171}{4} \lambda^2 -54\frac{\lambda}{\rho} >0.
   \end{align}
    Here, we have only kept the leading terms in the definition of $a_i$ ($i=0-3$) for simplicity. It appears  that the coefficients $a_i$ ($i=0-4$) of Eq. \eqref{eq-omega} are all positive definite. Consequently, Eq. \eqref{eq-omega} admits only negative roots $\omega <0$, meaning that the anisotropic fixed point is indeed stable against the perturbations. More interestingly, the numerical calculations shown in the Fig. \ref{fig1} indicate that the anisotropic fixed point is indeed an attractor solution to the dynamical system. These results, both analytical and numerical, strongly imply the violation of the cosmic no-hair conjecture in the present model.
\section{Tensor-to-scalar ratio} \label{sec5}
For heuristic reasons, we would like to investigate the corresponding tensor-to-scalar (power) ratio of the anisotropic power-law $k$-inflation. This can be easily done due to our previous works on the CMB imprints of non-canonical anisotropic inflation, partially done in Ref. \cite{SDBI} and fully done in Ref. \cite{Do:2020ler}. It is worth noting that recent observational constraints of isotropic power-law inflation has been investigated in Ref. \cite{Unnikrishnan:2013vga}.

As mentioned above, once the statistical isotropy of CMB is broken, the scalar power spectrum, i.e. the $TT$ correlations, will be modified accordingly as pointed out in Ref. \cite{ACW} as
\begin{equation} \label{ACW formula}
{\cal P}^\zeta_{k,{\text {ani}}}  = {\cal P}^{\zeta(0)}_{k} \left(1+g_\ast \cos^2 \theta_{{\bf k},{\bf V}}\right) .
\end{equation} 
Here $g_\ast$  is a constant expected to be smaller than one, i.e., $|g_\ast| <1$ since it characterizes  the deviation from the spatial isotropic. Hence, $|g_\ast \cos^2 \theta_{{\bf k},{\bf V}}| < 1$ is regarded as a small correction to the scalar power spectrum. In addition,  $\theta_{{\bf k},{\bf V}}$ is the angle between the comoving wave number $\bf k$ with the {\it privileged} direction $\bf V$  close to the ecliptic poles \cite{ACW}.  For convenience, we will write $\theta_{{\bf k},{\bf V}}$ as $\theta$ from now on. In addition, ${\cal P}^{\zeta(0)}_{k} $  is the isotropic scalar power spectrum, whose general definition for non-canonical scalar fields is given by \cite{k-inflation}
\begin{equation}  \label{def of P-zeta-2}
{\cal P}^{\zeta(0)}_{k}  = {\cal P}^{\zeta(0)}_{k,{\text{nc}}} = \frac{1}{8\pi^2 M_p^2}\frac{H^2}{c_s \epsilon},
\end{equation}
where $\epsilon\equiv -\dot H/H^2$ is the slow-roll parameter with $H$ being the Hubble expansion rate. Additionally, $c_s$ is understood as the speed of sound defined by \cite{k-inflation}
\begin{equation} \label{speed of sound}
c_s^2 \equiv \frac{\partial_X p}{\partial_X \rho}=\frac{\rho+p}{2X \partial_X \rho},
\end{equation}
with $p$ and $\rho$  the pressure and energy density parameters defined as
\begin{align}
p &=   P(\phi,X), \\
\rho &=   2X \partial_X P(\phi,X)-P(\phi,X),
\end{align}
respectively. 
According to Refs. \cite{SDBI,Do:2020ler}, the general scalar and tensor power spectra for a wide class of anisotropic $k$-inflation are given by 
\begin{equation} \label{general anisotropic scalar power spectrum}
{\cal P}^\zeta_{k,{\text{nc}}}  
 =  {\cal P}^{\zeta(0)}_{k,{\text{nc}}} \left(1- c_s^5 g_\ast^{0}  \sin^2 \theta \right)
\end{equation}
and
\begin{equation} 
{\cal P}^h_{{k,\text{nc}}}  
\simeq   {\cal P}^{h(0)}_{k,\text{nc}} \left(1- \frac{ \epsilon g_\ast^{0}}{4} \sin^2 \theta  \right),
\end{equation}
respectively. In addition, ${\cal P}^{h(0)}_{k,\text{nc}}$ is the isotropic tensor power spectrum for non-canonical scalar field defined as \cite{k-inflation}
\begin{equation} \label{def.of.Ph}
{\cal P}^{h(0)}_{k,\text{nc}}=  16 c_s \epsilon {\cal P}^{\zeta(0)}_{k,\text{nc}}.
\end{equation}
Consequently, a general value of $g_{\ast}$ for non-canonical anisotropic inflation can be figured out such as
\begin{equation} \label{definition_of_g}
g_{\ast}= g_{\ast}^{\rm nc} =  c_s^5 g_\ast^0 ,
\end{equation}
with $g_\ast^0$, whose definition has been shown in Refs. \cite{SDBI,Do:2020ler},  is for the canonical scalar field, i.e. $g_\ast=g_\ast^0$ when $c_s=1$. As a result, the corresponding tensor-to-scalar ratio of non-canonical anisotropic inflation turns out to be \cite{SDBI,Do:2020ler}
\begin{equation}  \label{general r-1}
r_{\rm nc} = \frac{{\cal P}^h_{{k,\text{nc}}}}{{\cal P}^\zeta_{k,{\text{nc}}} } = r_{\rm nc}^{\rm iso} \frac{6-\epsilon g_\ast^{0}}{6-4c_s^5 g_\ast^{0}},
\end{equation}
where  $r_{\rm nc}^{\rm iso} \equiv 16c_s\epsilon$ is the well-known tensor-to-scalar ratio for isotropic $k$-inflation \cite{k-inflation}, while $N_{c_sk} $ is the e-fold number, which is usually taken to be 60. 

As a result, the corresponding general scalar and tensor spectral indices of non-canonical anisotropic  inflation can be shown to be \cite{SDBI,Do:2020ler}
\begin{align} 
 n_s - 1 \simeq &  -2\epsilon - \tilde \eta -s +\left(\frac{2}{N_{c_sk}}-5s\right) \frac{2  c_s^5 g_\ast^0 }{3-2c_s^5 g_\ast^0  }, \\
 n_t \simeq &-2\epsilon,
\end{align}
respectively, where $\tilde \eta \equiv \dot\epsilon/(\epsilon H)$ and $s\equiv \dot c_s/(c_s H)$ \cite{k-inflation}.

Given these general expressions for a wide class of non-canonical anisotropic inflation, we now consider the present $k$-inflation model with 
\begin{equation}
P(\phi,X)= k_0 X + l_0 \exp \left[\lambda \phi \right] X^2.
\end{equation}
As a result, it is straightforward to define the corresponding speed of sound as 
\begin{equation}
c_s^2 = \frac{k_0+2l_0 \exp[\lambda \phi] X}{k_0+6l_0 \exp[\lambda \phi] X}.
\end{equation}
Thanks to the anisotropic power-law inflation shown above, it is straightforward to define the corresponding  $c_s^2$ to be
\begin{equation}
c_s^2 =\frac{k_0 +\xi^2 u}{k_0 +3 \xi^2 u} \simeq -\frac{\lambda}{4 \rho} \simeq \frac{1}{4 \zeta} \ll 1
\end{equation}
along with the slow-roll parameter 
\begin{equation} \label{epsilon-cs-2}
\epsilon = \frac{1}{\zeta} \simeq 4c_s^2 \ll 1.
\end{equation}
In addition, it turns out that $\tilde \eta = s=0$ for the current solution, which leads to
\begin{equation} \label{ns}
n_s-1 \simeq -2\epsilon+ \frac{4  c_s^5 g_\ast^0 }{N_{c_sk} \left(3-2c_s^5 g_\ast^0  \right)}.
\end{equation}

Note that the observational bounds of $g_\ast^0$ for the KSW models of canonical scalar field  have been investigated  in the light of the WMAP and Planck \cite{data}. In particular, an analysis has been done to give  another bound as  $|g_\ast^0|<0.072$  at 95\% confidence level (CL)  using the 9-year WMAP data. More recently, a general analysis using the Planck 2015 data has been performed to give the bounds of $g_\ast^0$ that $-0.041<g_\ast^0 <0.036$ at 95\% CL \cite{data}.  

\begin{figure}[hbtp] 
\begin{center}
{\includegraphics[height=80mm]{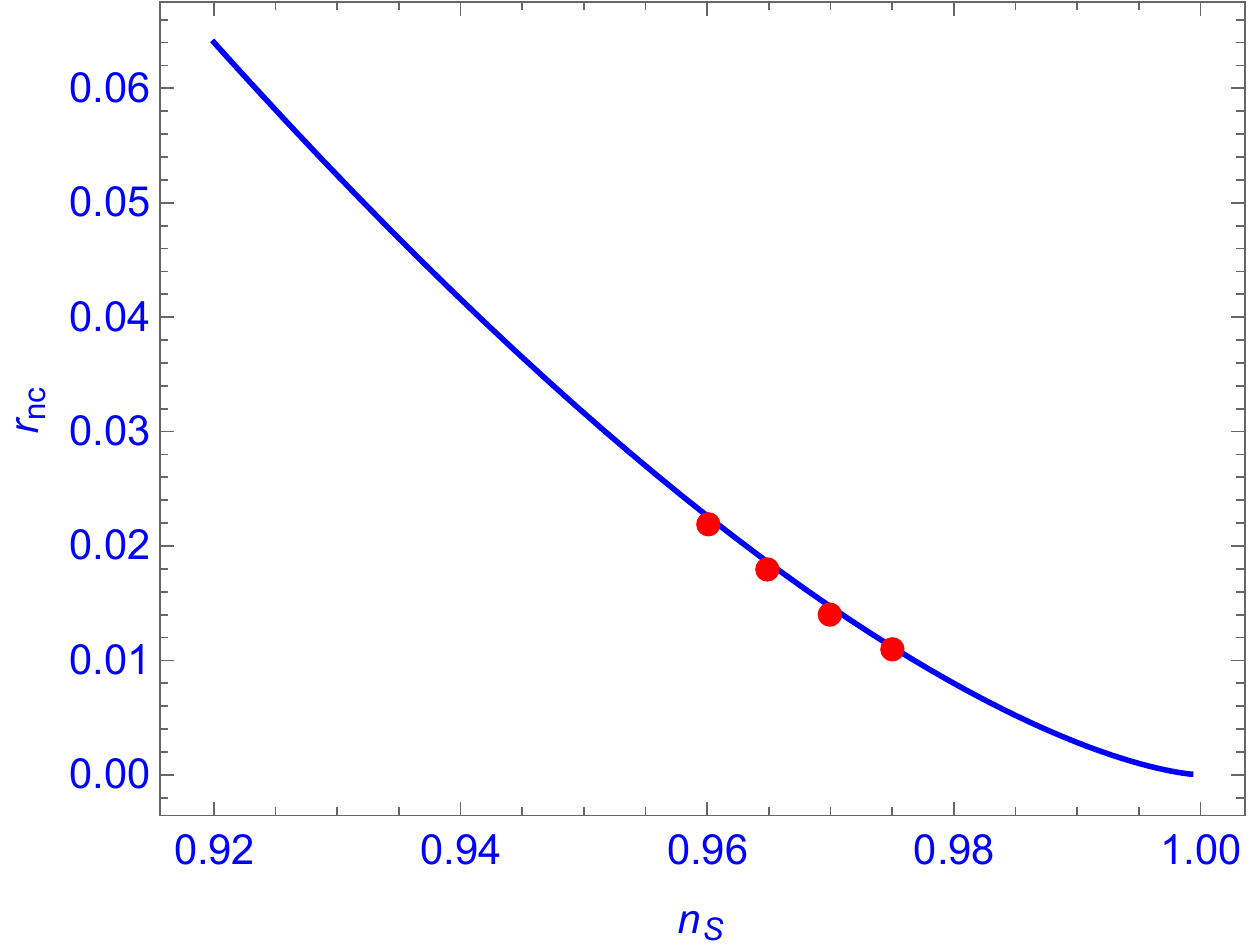}}\\
\caption{The $n_s - r_{\rm nc}$ diagram for the anisotropic power-law inflation found in this paper with $g_\ast^0= -0.03$ and $10^{-2}\leq c_s \leq 10^{-1}$. Four red points have the corresponding values such as $(n_s,r_{\rm nc})\simeq (0.96,0.022)$ for $c_s\simeq0.07$, $(0.965,0.018)$ for $c_s\simeq0.066$, $(0.97,0.014)$ for $c_s\simeq0.06$, and $(0.975,0.011)$ for $c_s\simeq0.056$. }
\label{fig2}
\end{center}
\end{figure}

For a comparison with the observational data of the Planck 2018 \cite{Planck}, we would like to plot the $n_s - r_{\rm nc}$ diagram using Eqs. \eqref{general r-1} and \eqref{ns} with the help of Eq. \eqref{epsilon-cs-2}. In addition, we will assume that $g_\ast^0= -0.03$  and a range of the speed of sound will  be limited as $10^{-2}\leq c_s \leq 10^{-1}$. According to the Fig. \ref{fig2}, especially its focused region where $0.96 \leq n_s \leq 0.975$, which overlaps with the most CL region for $n_s$ and $r$ of the Planck 2018, it turns out that our present model is highly consistent with the data of the Planck 2018. To end this section, we would like to note again that other CMB imprints of the general non-canonical anisotropic inflation, such as the $TB$ and $EB$ correlations, which vanish in isotropic inflation but not in anisotropic inflation, have been investigated in our previous work \cite{Do:2020ler}.
\section{Conclusions}\label{final}
We have shown that the non-canonical extension of the KSW model, in which the canonical scalar field has been replaced by the non-canonical field of the so-called {\it k}-inflation \cite{k-inflation}, does admit the  Bianchi type I spacetime as its stable and attractor solution during the inflationary phase. This study together with the previous ones done in Refs. \cite{MW,WFK1,ghost-condensed,SDBI,G} indicate that the  cosmic no-hair conjecture proposed by Hawking and his colleagues \cite{GH}  is extensively broken down in the KSW and its non-canonical extensions due to the existence of the unusual coupling between the scalar and electromagnetic field $-f^2(\phi)F_{\mu\nu}F^{\mu\nu}/4$. It is noted again that  the CMB imprints of non-canonical anisotropic inflation have been investigated in Ref. \cite{Do:2020ler}. Once these imprints  were confirmed by more sensitive primordial gravitational wave observations, then the present paper would provide one more non-canonical anisotropic inflation scenario, which might be useful to figure out the most viable anisotropic inflation model. For heuristic reasons,  we have investigated, based on the general results derived for a wide class of non-canonical anisotropic inflation in Refs. \cite{SDBI,Do:2020ler}, the corresponding tensor-to-scalar power ratio of this model. As a result, it has been shown to be highly consistent with the observational data of the Planck 2018. We hope that our present study would shed more light on the cosmological implications of the KSW anisotropic inflation as well as its non-canonical extensions.
\begin{acknowledgments}
The author would like to thank referees very much for their comments and suggestions, which are very useful to improve this paper.  The author would also like to thank Prof. W. F. Kao as well as Dr. Ing-Chen Lin very much for their fruitful collaborations on the previous works of  anisotropic inflation. This study is supported by the Vietnam National Foundation for Science and Technology Development (NAFOSTED) under grant number 103.01-2020.15.
\end{acknowledgments}
\appendix
\section{Field equations} \label{app1}
As a result, the following non-vanishing $00$, $11$, and {$22$ ($33$)} components of the Einstein field equations \eqref{Einstein} can be defined to be
\begin{align} \label{00}
3 \left(\dot\alpha^2 -\dot\sigma^2\right) =&~\frac{K}{2} \dot\phi^2 +\frac{3L}{4}\dot\phi^4+\frac{f^{-2}}{2}\exp\left[4\alpha -4\sigma\right] p_A^2, \\
\label{11}
2\left(\ddot\alpha +\ddot\sigma \right)+3 \left(\dot\alpha+\dot\sigma\right)^2 =& -\frac{K}{2}\dot\phi^2 -\frac{L}{4}\dot\phi^4 +\frac{f^{-2}}{2}\exp\left[4\alpha -4\sigma\right] p_A^2 , \\
\label{33}
 2\ddot\alpha -\ddot\sigma +3\left(\dot\alpha^2-\dot\alpha \dot\sigma+\dot\sigma^2\right) =&-\frac{K}{2}\dot\phi^2 -\frac{L}{4}\dot\phi^4 -\frac{f^{-2}}{2}\exp\left[4\alpha -4\sigma\right] p_A^2,
\end{align}
respectively. It is clear that $00$-component equation \eqref{00} is identical to Eq. \eqref{field-eq-1}, which is called the Friedmann equation.  As a result, eliminating $\ddot\alpha$ in both Eqs. \eqref{11} and \eqref{33} leads to the anisotropy equation  \eqref{field-eq-3}. On the other hand, eliminating $\ddot\sigma$ in both Eqs. \eqref{11} and \eqref{33}  leads to Eq. \eqref{field-eq-2} with the help of the Friedmann equation \eqref{00}.

\end{document}